\documentclass[epj]{svjour}
\usepackage{graphics,amsmath,amssymb,multirow,color,pstricks,pst-plot}
\usepackage{xspace,graphicx,comment,hyperref,cite}

\newcommand{\QQbar}{\ensuremath{Q\overline{Q}}\xspace}
\newcommand{\pt}{\ensuremath{p_{\rm T}}\xspace}

\newcommand{\jpsi}{\ensuremath{{\rm J}/\psi}\xspace}
\newcommand{\psip}{\ensuremath{\psi{\rm (2S)}}\xspace}

\newcommand{\oneSzero}{\ensuremath{{^1{\rm S}_0^{[8]}}}\xspace}
\newcommand{\threeSone}{\ensuremath{{^3{\rm S}_1^{[8]}}}\xspace}
\newcommand{\threePJ}{\ensuremath{{^3{\rm P}_J^{[8]}}}\xspace}
\newcommand{\lincomb}{\ensuremath{{\threeSone + \kappa \, \threePJ}}\xspace}

\newcommand{\threePoneSinglet}{\ensuremath{{^3{\rm P}_1^{[1]}}}\xspace}
\newcommand{\threePtwoSinglet}{\ensuremath{{^3{\rm P}_2^{[1]}}}\xspace}
\newcommand{\threeSoneSinglet}{\ensuremath{{^3{\rm S}_1^{[1]}}}\xspace}

\newcommand{\ptM}{\ensuremath{p_{\rm T}/M}\xspace}
\newcommand{\lth}{\ensuremath{\lambda_\vartheta}\xspace}
\newcommand{\lph}{\ensuremath{\lambda_\varphi}\xspace}
\newcommand{\ltp}{\ensuremath{\lambda_{\vartheta\varphi}}\xspace}

\begin{document}

\title{NRQCD colour-octet expansion vs.\ LHC quarkonium production:\\
signs of a hierarchy puzzle?}

\author{Pietro Faccioli\inst{1} 
\and Carlos Louren\c{c}o\inst{2}
}
\institute{LIP, Lisbon, Portugal
\and 
CERN, Geneva, Switzerland}

\date{Received: April 27, 2019 / Revised version: date}
\abstract{
The observation of unpolarized quarkonium production in high energy pp collisions, at mid rapidity,
implies a significant violation of the non-relativistic QCD (NRQCD) velocity scaling rules. 
A precise experimental confirmation of this picture could definitely rule out the current formulation of the factorization expansion. 
This conclusion relies on current perturbative determinations of the short-distance kinematic factors 
and may be reverted if improved calculations would modify, in a very specific way, their transverse momentum dependences. 
That solution would result, however, in a full degeneracy in the presently assumed basis of $^{2S+1}L_J$ Fock states.
Therefore, whatever the outcome, 
improved polarization measurements will challenge and improve our fundamental understanding of quarkonium production.
}

\titlerunning{NRQCD colour-octet expansion vs.\ LHC quarkonium production}

\maketitle

Quarkonium production is a central case study for the understanding of QCD bound state formation. 
Non-relativistic QCD (NRQCD) addresses its description in a rigorous way starting from first principles, 
assuming the factorization of short- and long-distance effects in the limit of small relative velocity ($v$) 
of the heavy quark and antiquark (\QQbar) forming the bound state~\cite{bib:NRQCD}. 
Under this hypothesis, the \QQbar state is modelled mathematically as an expansion over Fock states
of determined angular momentum and colour properties, $\QQbar(^{2S+1}L_J^{[n]})$, 
with $L = S, P,$ etc., $S = 0, 1,$ etc., $J = 0, 1, 2,$ etc., and $n = 1$ (colour singlet) or 8 (colour octet).
The non-perturbative evolution of such pre-resonance states to the observed quarkonium is described by constant factors 
(long-distance matrix elements, LDMEs, $\mathcal{L}$), 
currently not calculable and determined in global fits to sets of measurements, 
while perturbative calculations fix the kinematics-dependent short-distance \QQbar production cross sections 
(short distance coefficients, SDCs, $\mathcal{S}$) 
for each of the considered terms of the expansion,
so that the production cross section of the quarkonium state $H$ in pp collisions is
\begin{align}
\label{eq:factorization}
\nonumber
\begin{split}
& \sigma({\rm pp} \to H+X) = \\
& \sum\limits_{S,L,n} \mathcal{S}\Big({\rm pp} \to \QQbar[^{2S+1}L_J^{[n]}] + X; \sqrt{s}, M, \pt, y\Big) \\
& \times \mathcal{L}\Big(\QQbar[^{2S+1}L_J^{[n]}] \to H\Big)  \; ,
\end{split}
\end{align}
where $M$, \pt, and $y$ are the mass, transverse momentum, and rapidity of the \QQbar.


After a decades-long inconsistency between theory predictions and measurements, 
concerning in particular the polarizations, 
a major progress was brought by detailed LHC measurements 
of cross sections~\cite{bib:ATLASdimuon,bib:ATLASpsi2S,bib:ATLASchic,bib:ATLASYnS,bib:CMSjpsi,bib:CMSYnS,bib:LHCb_psi_cs,bib:LHCb_psip_cs}
and polarizations~\cite{bib:CMSlambdaPsi2S,bib:CMSlambdaYnS,bib:LHCb_psi_pol,bib:LHCb_psip_pol},
better calculations~\cite{bib:BKMPLA,bib:Ma:2010jj,bib:Chao:2012iv,bib:Shao:2014fca,bib:Shao:2015vga,bib:BodwinCorrections,bib:Bodwin:2015iua},
and improved methods for unbiased data-theory comparisons~\cite{bib:EPJC69,bib:FaccioliPLB736,bib:Faccioli:PLB773}.

Existing NLO SDC calculations can quantitatively describe the observed ``universal unpolarized" scenario of LHC mid-rapidity data, 
where all states are seemingly produced with identical \ptM distributions and zero polarizations~\cite{bib:Faccioli:EPJC78p268}. 
From a conceptual point of view, however, the structure of the NRQCD factorization expansion, 
where three kinematically very different octet terms (\oneSzero, \threeSone, and \threePJ) dominate the production of S-wave states 
and two singlet terms (\threePoneSinglet and \threePtwoSinglet) are additionally necessary 
(summed to the \threeSone term) to describe $\chi_{c1,b1}$ and $\chi_{c2,b2}$ production,
does not seem to naturally match the exceptionally simple patterns seen in the experimental data. 
In fact, the theory accomplishes its formal success thanks to unexpected, precise cancellations.

While the seemingly superfluous complexity of the theory formulation is particularly apparent in
the comparison with the remarkably simple experimental patterns,
a certain degree of theory redundancy can be seen independently of any experimental observation.
For S-wave quarkonium production, which is the focus of this paper,
the dominating terms of the NRQCD expansion are the \oneSzero, \threeSone, and \threePJ octets, all of the same order, $v^4$
(the \threeSoneSinglet singlet term being negligible because of its small SDC). 
Throughout this paper we use the next-to-leading order (NLO) calculations of Ref.~\cite{bib:Shao:2014fca,bib:Shao:2015vga}
for the SDCs as functions of \pt, in pp collisions at $\sqrt{s} = 7$\,TeV and mid rapidity. 

\begin{figure}[t]
\centering
\includegraphics[width=1.0\linewidth]{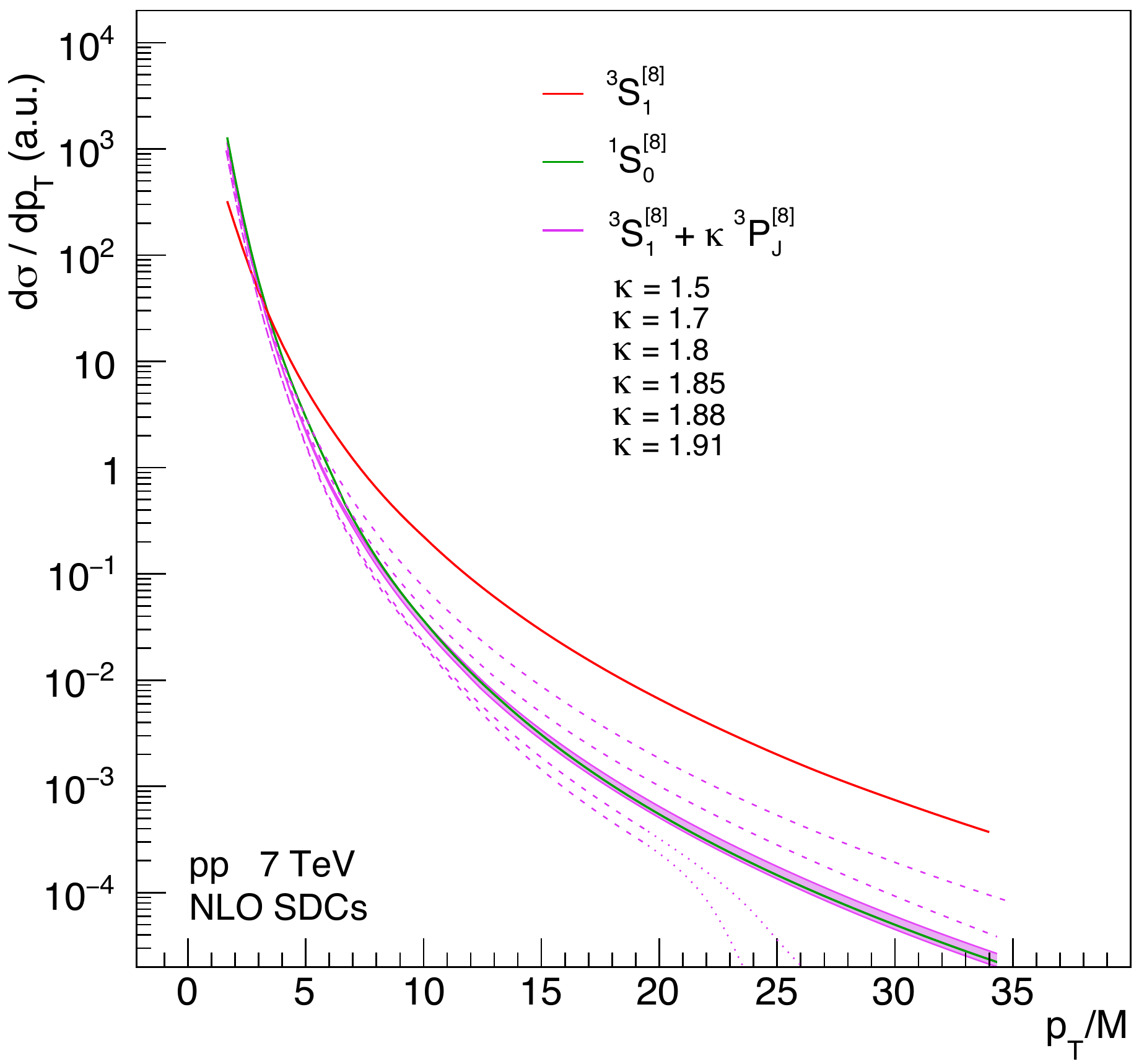}
\caption{The dominating colour-octet SDCs for the direct production of $^3{\rm S}_1$ quarkonia in NLO NRQCD, as functions of \ptM:
\oneSzero (green), \threeSone (red), and several \lincomb combinations (magenta). 
The band represents the $\kappa = 1.8$--1.85 range, which matches very well the shape of the (arbitrarily normalized) \oneSzero SDC.}
\label{fig:theory_SDCs}
\end{figure}

\begin{figure}[t]
\centering
\includegraphics[width=1.0\linewidth]{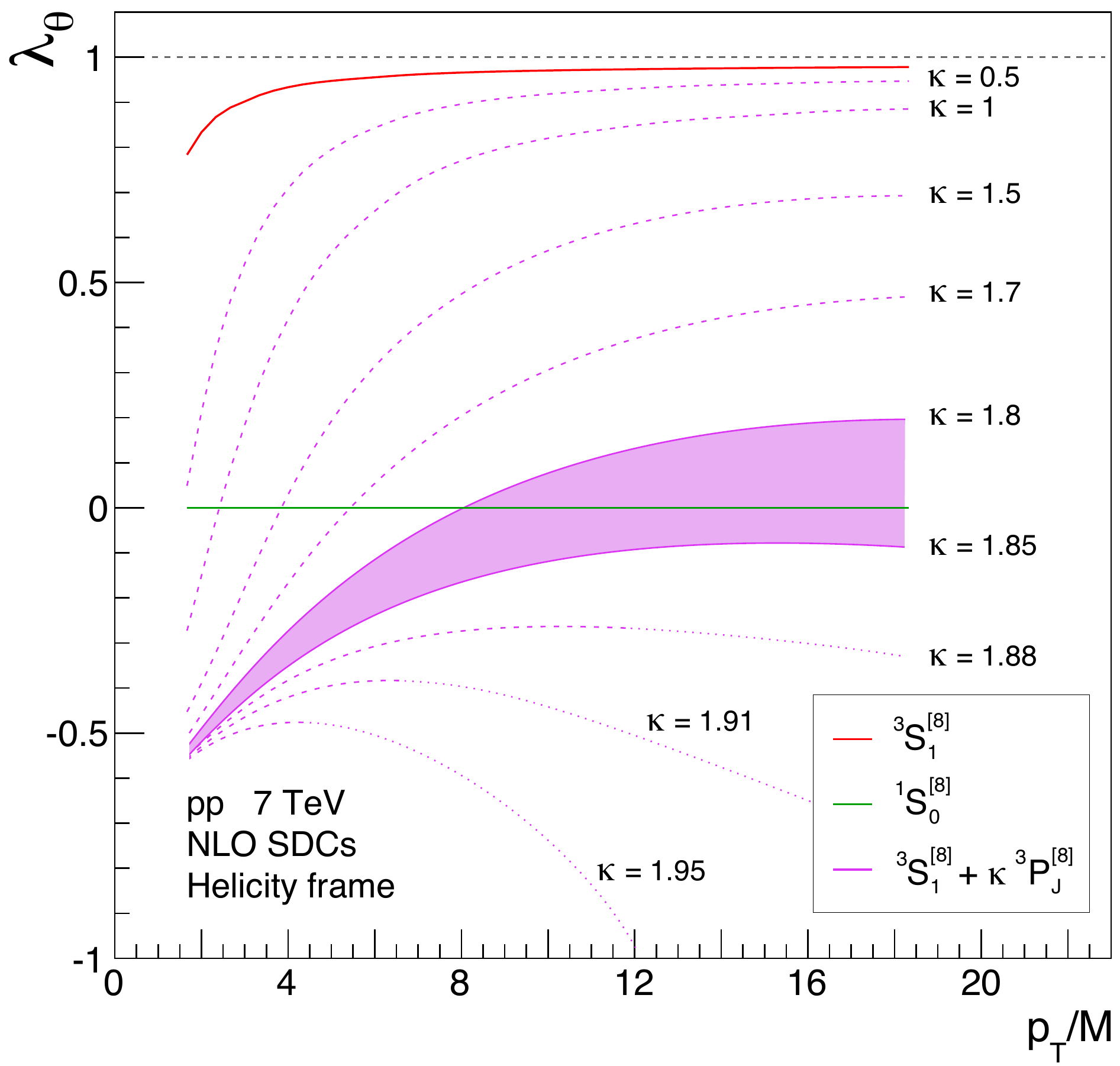}
\caption{The polarization parameter \lth for the different components of $^3{\rm S}_1$ quarkonium production in NLO NRQCD, as functions of \ptM:
\oneSzero (green), \threeSone (red), and several \lincomb combinations (magenta). 
The band represents the $\kappa = 1.8$--1.85 range, as in Fig.~\ref{fig:theory_SDCs}.}
\label{fig:theory_pol}
\end{figure}

Figure~\ref{fig:theory_SDCs} shows that the shape of the \oneSzero SDC is indistinguishable from a linear combination of the two other terms,
$\threeSone + \kappa \, \threePJ$, with $\kappa$ between 1.8 and 1.85. 
While in principle one would expect to fit experimental data with a superposition of three independent terms,
leading to a definite determination of the three corresponding LDMEs,
it turns out that the presently-available NLO SDCs are not \emph{independent} kinematic templates 
and even very precise measurements of \pt-differential cross sections will only be able to determine, 
in the best case, two parameters (as already noted in Ref.~\cite{bib:Ma:2010jj}).
In other words, the NLO description of the \pt dependence of quarkonium production has three theory parameters, 
but only two degrees of freedom.
This observation suggests 
the existence of a spurious element of complexity in the assumed base of subprocesses.

According to the present NLO knowledge of the SDCs, the polarization is a better discriminating observable.
Figure~\ref{fig:theory_pol} compares the calculated dilepton-decay polar anisotropy parameter, \lth, as a function of \ptM, 
for the \oneSzero, \threeSone, and \lincomb terms.
Contrary to the case of the production yields, 
no value of $\kappa$ exists for which $\lth(\threeSone + \kappa \threePJ) \approx \lth(\oneSzero)$ 
as a function of \ptM, 
and a momentum-dependent polarization measurement should be able to disentangle the three components. 
However, for $\ptM \gtrsim 10$, the \lincomb combination with $\kappa$ in the 1.8--1.85 range becomes unpolarized, just as the \oneSzero term.
Therefore, at high \pt, the polarization measurements are also unable to resolve more than two degrees of freedom in the space of the contributing LDMEs,
at least for quarkonia produced at mid-rapidity in high-energy pp collisions.

It is worth noting that for $\kappa \gtrsim 1.85$ the \lincomb combination assumes unphysical behaviours,
represented in Figs.~\ref{fig:theory_SDCs} and~\ref{fig:theory_pol} by the dotted lines: 
the \ptM distribution shows seemingly anomalous changes of curvature at high \ptM 
while the polarization parameter starts decreasing towards unphysical asymptotic values ($\lth < -1$).
Intriguingly, the ``degeneracy condition'' $\kappa = 1.8$--1.85 happens just before the border of the physical domain of positivity.

\begin{figure}[t]
\centering
\includegraphics[width=1.0\linewidth]{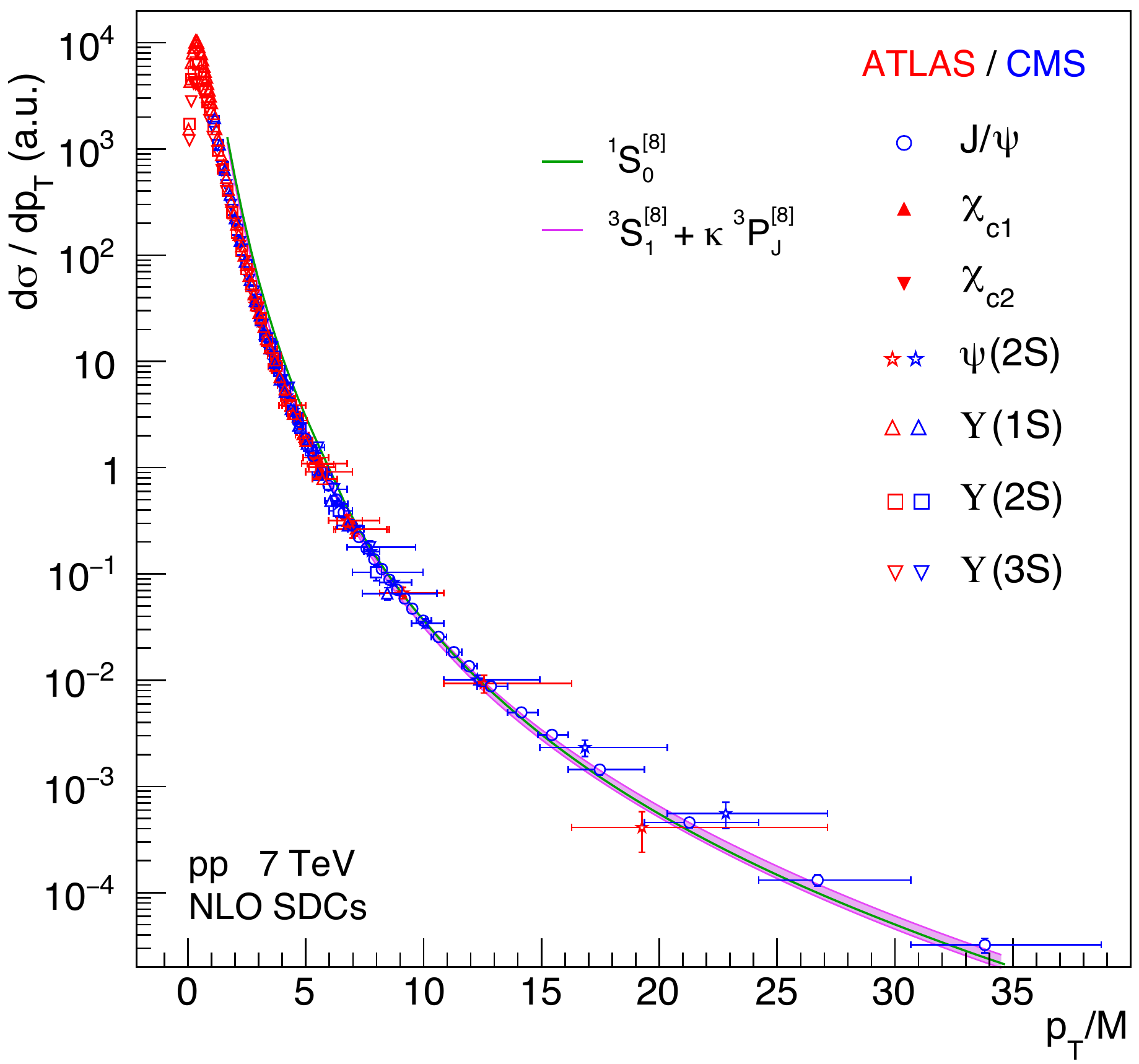}
\caption{Mid-rapidity quarkonium cross sections measured in pp collisions at $\sqrt{s} = 7$\,TeV 
by ATLAS (red markers)~\cite{bib:ATLASpsi2S, bib:ATLASYnS, bib:ATLASchic}
and CMS (blue markers)~\cite{bib:CMSjpsi, bib:CMSYnS}, 
with normalizations arbitrarily adjusted to the \jpsi points to illustrate the universality of the \ptM dependence.
The data are compared to the shapes of the \oneSzero octet (green) 
and of the \lincomb combination with $\kappa = 1.8$--1.85 (magenta band).}
\label{fig:data_vs_SDCs}
\end{figure}

\begin{figure}[t]
\centering
\includegraphics[width=1.0\linewidth]{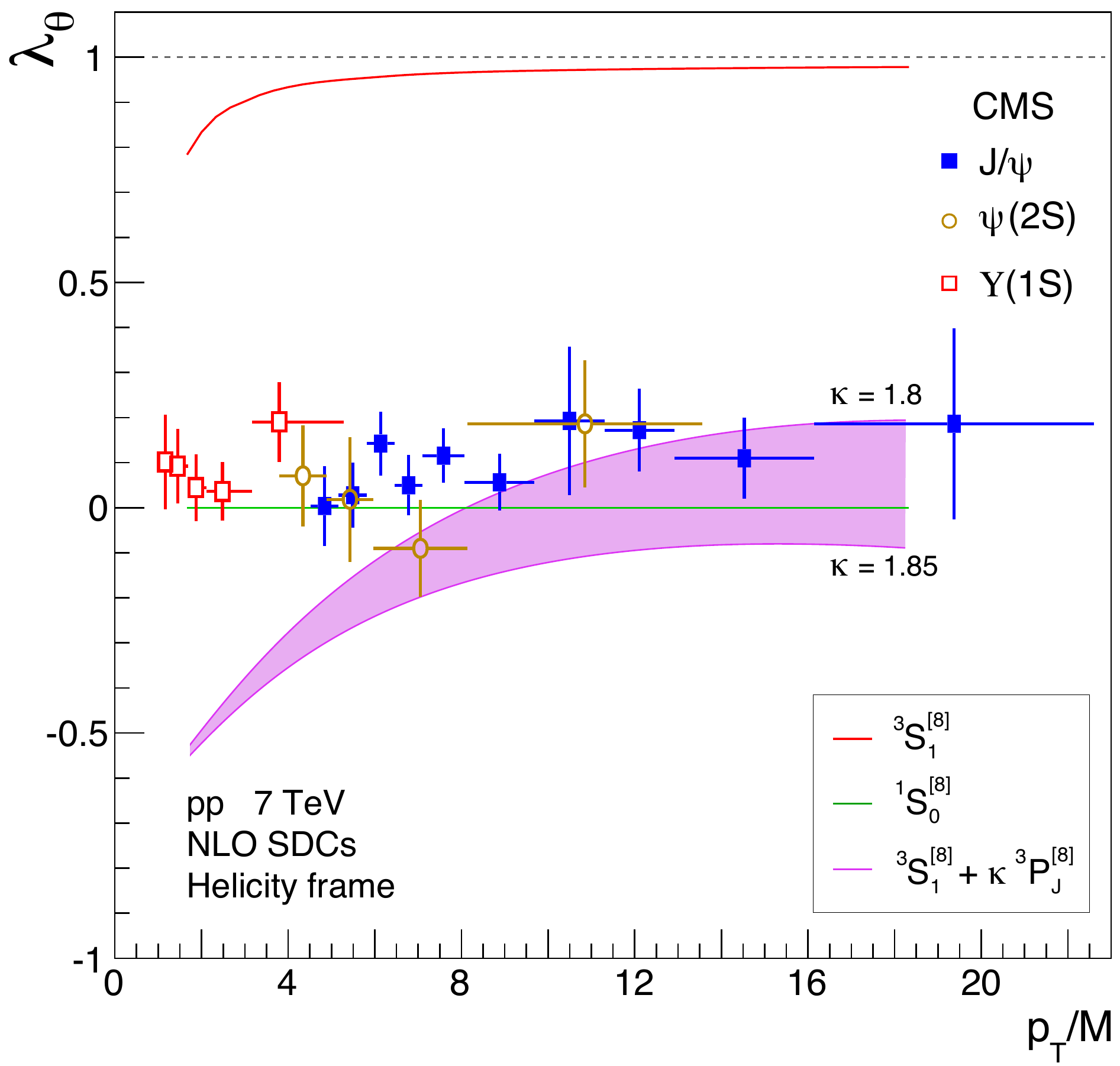}
\caption{The \jpsi, \psip, and $\Upsilon$(1S) polarization parameters as a function of \ptM, 
measured at mid-rapidity by the CMS experiment, compared to the NRQCD calculations for the 
\oneSzero (green) and \threeSone (red) octets, 
as well as to the \lincomb combination with $\kappa = 1.8$--1.85 (magenta band).}
\label{fig:data_vs_theory_pol}
\end{figure}

We will now see what the experimental measurements can add to the picture.
Figures~\ref{fig:data_vs_SDCs} and~\ref{fig:data_vs_theory_pol} show 
the mid-rapidity LHC measurements of cross sections and polarizations of different charmonium and bottomonium states.

As already mentioned, no significant differences are observed in the \ptM dependences of yields and polarizations 
for the different states, 
despite their varying feed-down contributions (ranging between 0 and 40\%) from heavier quarkonia. 
This universal behaviour has been discussed in detail in Ref.~\cite{bib:Faccioli:EPJC78p268},
where it is shown that the $\chi_c$ feed-down contributions do not alter significantly, 
within the current experimental precision, the kinematic patterns of the directly-produced \jpsi mesons. 
It is, therefore, reasonable to compare the inclusive data (including feed-down) 
to the NRQCD curves computed for direct production (excluding feed-down). 
For the curves themselves we use the NLO SDC calculations~\cite{bib:Shao:2014fca,bib:Shao:2015vga} 
performed for a \QQbar mass of 3\,GeV, 
assuming the validity of the \ptM scaling observed in cross section data as a general rule for individual processes. 
This approximation is justified by dimensional analysis reasonings 
applied to a single production mechanism~\cite{bib:Faccioli:EPJC78p118},
when $\sqrt{s}$ is large with respect to the momentum and mass of the observed state.

The figure shows that the scenario $\kappa = 1.8$--1.85 is not just an hypothetical idea to be considered among many others
but a scenario in rather good agreement with the presently existing quarkonium cross section and polarization measurements.
For $\ptM \gtrsim 10$, the cross section measurements are compatible with the \ptM distribution of the \oneSzero term, 
only a small correction coming from other terms. 
The observed lack of polarization further supports the idea that the \oneSzero term is close to reproducing the data 
with no need of further contributions. 
For high \ptM, the NRQCD expansion appears, therefore, as fully degenerate: 
all the data can be described by a single one of the basis terms, the \oneSzero octet, 
or, equivalently, by the combination \lincomb, with $\kappa = 1.8$--1.85.

As previously mentioned, the degeneracy is broken for $\ptM < 10$, where the NLO SDCs lead to 
different polarization predictions.
However, the \emph{measured} polarizations remain independent of \ptM down to $\ptM \sim 1$.
This means that the comparison between the measured data and the NLO SDCs 
clearly excludes a significant contribution of the \lincomb term, 
which is expected to be either strongly \pt dependent or strongly polarized,
as shown in Fig.~\ref{fig:theory_pol}. 
Even a relatively small \lincomb contribution to the quarkonium production yields, of around 10\%,
is already excluded by the presently existing polarization data, independently of the value of $\kappa$.
In other words, the mid-rapidity quarkonium polarization measurements made at the LHC
indicate a strong hierarchy of cross section contributions,
clearly enhancing the \oneSzero octet 
and suppressing the \emph{individual} \threeSone and \threePJ terms,
not only their partially cancelling combination.
This observation constitutes a significant violation of the $v$-scaling rules, 
according to which the three LDMEs should have the same order of magnitude.

Such a strong and unexpected constraint relies on the accuracy of the present NLO calculations.
Uncertainties from the factorization scale and from the quark masses, 
mainly affecting the SDC normalizations, 
have a negligible impact on our considerations, 
which are entirely based on shape differences.
Since the shapes of the NLO \oneSzero and \threeSone SDCs are similar to the leading order (LO) versions,
it is reasonable to expect that future computations of higher-order contributions will not reveal significant changes, 
as indicated by the already known partial corrections 
due to fragmentation contributions~\cite{bib:BodwinCorrections,bib:Bodwin:2015iua}.
The \threePJ term, instead, shows drastic changes from LO to NLO, including a change of sign,
and the fragmentation corrections are large~\cite{bib:BodwinCorrections,bib:Bodwin:2015iua}.
One may argue, hence, that higher-order corrections could change the \ptM dependence of the \lincomb polarization, 
especially towards low \ptM. 
To reach a situation where the polarization measurements would no longer exclude this contribution, 
so that the $v^2$ hierarchy predicted by NRQCD (\oneSzero $\sim$ \threeSone $\sim$ \threePJ) would be recovered,
those future higher-order corrections should necessarily lead to a flatter dependence of \lth on \ptM.
The presently available quarkonium polarization measurements are already sufficiently precise to suggest that
those future computations need to reach a suspiciously high level of fine tuning, 
but it is clear that new polarization measurements, of significantly improved precision, 
are needed to provide the ultimate constraint on the NRQCD hierarchy puzzle.

In particular, high-precision measurements confirming beyond doubt the scenario of vanishing and \pt-independent polarization 
would leave zero margin for a \lincomb contribution other than unpolarized and \pt-independent. 
The crucial importance of such measurements can be appreciated by considering the two possible conclusions: 
a)~if the calculated \lincomb term remains significantly different from unpolarized and shows some level of \pt dependence
(as is the case of the NLO computations), 
then we will be forced to conclude that its contribution is zero 
and quarkonium production is driven entirely by the \oneSzero channel; 
b)~if improved (higher order) calculations lead to $\lth(\threeSone + \kappa \threePJ) \sim 0$, 
the validity of the NRQCD $v$-scaling rules will be formally rescued,
but the three octet terms would collapse to a single one, 
given that they would then have identical observable kinematic properties.
This outcome would clearly point to the existence of a more natural and fundamental formulation 
of the factorization expansion, reflecting less degrees of freedom,
valid at least in the conditions of mid-rapidity quarkonium production in high-energy pp collisions.

The best way to accurately probe the low-\ptM region is to use $\Upsilon$(nS) measurements,
ideally with well-resolved 1S, 2S and 3S states
to indirectly examine possible effects from the feed-down decay contributions of the \mbox{P-wave} states. 
Accessing the high-\ptM kinematical domain 
(easier to study with \jpsi and \psip data, given the lower masses and higher production cross sections)
will be crucial to probe if quarkonium production 
is completely dominated by a single unpolarized production mechanism (the \oneSzero octet term, say) 
or if there are two (or more) polarized terms that seemingly cancel each other in the existing data. 
High-precision measurements revealing no polarization, independently of \pt, 
could no longer be explained with \emph{coincidental} cancellations between cross section terms: 
if the present hints of cancellations result from a conspiracy between the low experimental precision 
and the limited kinematical domain covered by data, 
a \pt-dependent residual polarization should become visible at sufficiently high \pt.
It should be easy to vastly reduce the statistical uncertainties of the 
presently available mid-rapidity LHC quarkonium polarization measurements, 
which are exclusively based on the 7\,TeV data collected by CMS in 2011, 
corresponding to an integrated luminosity of only 5\,fb$^{-1}$, 
much less than the 140\,fb$^{-1}$ collected at 13\,TeV between 2016 and 2018.
Polarization measurements reported by the LHCb collaboration~\cite{bib:LHCb_jpsi,bib:LHCb_psip,bib:LHCb_upsilon} 
will also be very valuable to precisely investigate the low \pt region, 
once the \ptM scaling studies~\cite{bib:Faccioli:EPJC78p268} 
will have been extended to the forward rapidity range covered by that experiment.
The systematic uncertainties, 
likely to become the dominating ones in a large range of the measured \pt spectrum, 
can be reduced by performing the measurements in at least two (orthogonal) polarization frames
and by also reporting measurements of the \lph and \ltp azimuthal anisotropy parameters~\cite{bib:EPJC69}.
Measuring the $\tilde{\lambda}$ frame-invariant parameter~\cite{bib:lambda_tilde}
in several polarization frames will also help uncovering potential biases.

It is important to note that the experimental conditions of the LHC data considered in this paper 
may represent our best chance of exploring the hierarchy and possible limit degeneracies 
in the NRQCD expansion. 
Indeed, given the very high collision energies, we can access mid-rapidity data up to relatively high \ptM values 
with a good statistical precision, so that we can study a phase space window 
where quarkonium production is dominated by $2 \to 2$ processes of the kind $gg \to \QQbar + g$. 
In this configuration the \QQbar pre-resonance can ``freely" assume all $J$ values up to, in principle, 
$J=3$ (with $S=0$ or $S=1$); there is, in principle, no \emph{a priori} topological constraint removing some of these terms, 
nor imposing a specific $J_z$ (i.e., a polarization). 
Instead, data obtained in experimental conditions dominated by specific individual channels 
would necessary be affected by additional constraints, 
as in the case of singlet-driven production via virtual photon in $e^+e^-$ collisions, 
or in kinematic domains where $2 \to 1$ production becomes important 
and the \QQbar directly inherits the angular momentum state of the system of colliding partons.
In short, it could well be that only data collected in optimal conditions, 
where no special angular momentum constraints are directly imposed on the produced \QQbar, 
allow us to probe the existence of an ``unbroken degeneracy" at a fundamental level. 

In summary, we have shown the existence of a kinematic domain in high-energy pp collisions, 
$\ptM \gtrsim 10$ at mid-rapidity,
where the description of quarkonium production using the current formulation of a $v^2$ expansion in $^{2S+1}L_J$ Fock-states, 
together with NLO SDC calculations, becomes degenerate.
Indeed, the observed lack of polarization and universal \ptM dependence of the yields
can be described, indifferently, by one of the expansion terms alone, the \oneSzero octet, 
or by a combination of the other two, the \lincomb linear combination
(and, obviously, by a mixture of the two cases, combined with unobservable proportions).
This degeneracy is broken for $\ptM < 10$ by the polarization prediction, 
because the \lincomb term acquires a strong \ptM dependence, 
differentiating it from the constant and unpolarized \oneSzero octet. 
However, such change of regime is not at all observed in the existing measurements, 
which seamlessly prolong their unpolarized behaviour down to the lowest data point, at $\ptM \sim 2$. 
This implies a strong violation of the $v$-scaling rules and an unexpected dominance of the \oneSzero term.
Future high-precision quarkonium polarization measurements 
have the potential to rule out the current formulation of the factorization expansion beyond any possible recovery, 
if they confirm the unpolarized scenario suggested by the currently available data. 
Fine-tuned higher-order corrections could provide SDC calculations that would resuscitate the \lincomb term,
but at the expense of turning it, necessarily, into another unpolarized term, 
thereby exposing a full degeneracy of the currently postulated Fock-space expansion. 
Such an outcome, rather than solving the riddle, would raise an even more compelling question: 
does a more essential and natural description of the experimental observations exist? 
The search for explanations of why and how the degeneracy appears in the considered phase space 
(mid-rapidity and/or high \ptM, in high-energy pp collisions), 
will certainly bring advances in the fundamental understanding of quarkonium production.  

\medskip

We would like to thank Hua-Sheng Shao, who kindly provided tables 
of the NLO calculations of the SDCs.

\bibliographystyle{cl_unsrt}
\bibliography{references}{}

\end{document}